\documentclass[fullpaper,twocolumn]{jpsj2}
\usepackage{txfonts}
\usepackage[dvipdfm]{color}
\usepackage{amssymb}
\usepackage{multirow}
\usepackage{amsmath}
\usepackage{wrapfig}
\usepackage{subfigure}

\newcommand{\be}{\begin{equation}}
\newcommand{\ee}{\end{equation}}
\newcommand{\bea}{\begin{eqnarray}}
\newcommand{\eea}{\end{eqnarray}}
\newcommand{\nn}{\nonumber \\}
\newcommand{\mr}{\mathrm}

\renewcommand{\figurename}{Fig.}

\title{Magnetic Chirality Induced from Ruderman-Kittel- Kasuya-Yosida Interaction at an Interface of a Ferromagnet/Heavy Metal Heterostructure
}

\author{Taira Shibuya, Hiroyasu Matsuura, and Masao Ogata
}
\inst{
Department of Physics, University of Tokyo, Bunkyo, Tokyo 113-0033, Japan
} 

\abst{
We study a microscopic derivation and the properties of the Dzyaloshinskii-Moriya interaction (DMI) between local magnetic moments in ferromagnet/heavy metal heterostructures. First, we derive DMI by Ruderman-Kittel-Kasuya-Yosida interaction through electrons in a heavy metal with Rashba spin orbit interaction (SOI). Next, we study the dependences of the DMI on the Rashba SOI, lattice constant, and chemical potential. We find that the DMI amplitude increases linearly when the Rashba SOI is small, has a maximum when the Rashba SOI is comparable to the hopping integral, and decreases when the Rashba SOI is large. The sign of the DMI  not only changes depending on the sign of the Rashba SOI but also the lattice constants and the chemical potential of the heavy metal. The implications of the obtained results for experiments are discussed.}

\begin{document}
\maketitle

\section{Introduction}
In a ferromagnet/heavy metal heterostructure, it has recently been reported that in-plane currents passing through the heavy metal switch the magnetization of the ferromagnet \cite{miron,liu1}. The system consists of an ultrathin ferromagnet with perpendicular anisotropy and a nonmagnetic heavy metal that has a large spin orbit interaction (SOI). In this system, it is argued that the electric current induces a spin torque via a spin Hall effect and a Rashba effect\cite{miron2, liu2, kimras, khval}. Then, the induced spin torque drives domain walls in the ferromagnet and, as a result, the magnetization is reversed. This current-induced domain wall motion in a ferromagnet/heavy metal heterostructure is attracting much interest as one of the key technologies of advanced memory devices in spintronics \cite{miron,liu1,miron2,liu2,kimras,khval, moore, miron3, suzuki ,emori,ryu}.

To investigate the mechanism of this current-induced domain wall motion, the Dzyaloshinskii-Moriya interaction \cite{dzyalo,mori} (DMI) between local magnetic moments near the interfaces has attracted considerable attention \cite{emori, ryu, heide, thiaville, kim}. The DMI has the form, $\mbox{\boldmath{$D$}}_{ij} \cdot (\mbox{\boldmath{$S$}}_i \times \mbox{\boldmath{$S$}}_j)$, where $\mbox{\boldmath{$D$}}_{ij}$ is the DMI vector and $\mbox{\boldmath{$S$}}_i$, $\mbox{\boldmath{$S$}}_j$ are magnetic moments on sites $i$, $j$, respectively. 
The amplitude of the DMI is of practical significance since the velocity of domain wall motion is roughly proportional to the amplitude of the DMI \cite{thiaville}. Furthermore, recent experiments have shown that the DMI in ferromagnet/heavy metal heterostructures can be controlled by engineering the interface \cite{chen, ryu2, torrejon}. It was found that not only the amplitude of the DMI but also its sign depends on the choice of the heavy metal. 

\begin{figure}[h]
\includegraphics[width=7cm,clip]{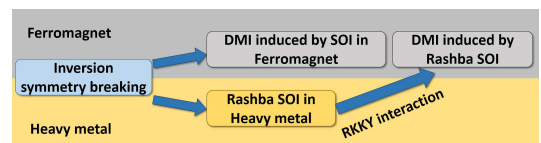}
\centering
\caption{(Color online) Schematic of the microscopic origins of DMI in a ferromagnet/heavy metal heterostructure. We consider the DMI induced by the Rashba SOI in the heavy metal layer through the RKKY interaction.}
\label{f00}
\end{figure}

In this paper, we consider microscopic origins of the DMI in a ferromagnet/heavy metal heterostructure as shown in Fig. \ref{f00} and investigate the heavy metal dependence of the DMI. There are found to be two mechanisms underlying the origin of the DMI. (1) Owing to the inversion symmetry breaking at the interface, DMI is induced by the SOI in the ferromagnet near the interface. It can be calculated by several methods. \cite{mori, kataoka}
(2) The other mechanism is that the Rashba SOI in the heavy metal induced by inversion symmetry breaking leads to DMI through the Ruderman-Kittel-Kasuya-Yosida (RKKY) interaction. Generally, the SOI in the ferromagnet is much smaller than that in the heavy metal. Therefore, we expect that mechanism (2) is dominant for understanding the heavy metal dependence of DMI. In the following, we consider mechanism (2).

Mechanism (2) has been studied by assuming a two-dimensional electron gas\cite{imamura}. 
However, the RKKY interaction is sensitive to the Fermi surface. Therefore, it is important to study the effect of the shape of the Fermi surface in the heavy metal on the DMI. In this study, we assume a tight-binding model in a two-dimensional square lattice for the heavy metal as a simplified model in order to extract the essence of the phenomena. 
We will discuss the dependences of DMI on the Rashba SOI, lattice constant, and chemical potential of the heavy metal. 

\section{Derivation of Equations}
\begin{figure}[h]
\includegraphics[width=7cm,clip]{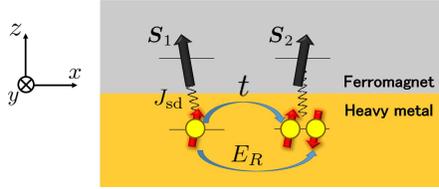}
\centering
\caption{(Color online) Schematic of the model. The hopping process $t$, the spin flipping process $E_R$ and the sd interaction $J_{\mr{sd}}$ are depicted.}
\label{f0}
\end{figure}

We derive the RKKY interaction from the second-order perturbation with respect to the sd interaction between a magnetic moment in the ferromagnet and an electron spin in the heavy metal.
The Hamiltonian is given by 
\begin{align}
\mathcal{H}&=\mathcal{H}_{\mr{hop}}+\mathcal{H}_{\mr{Rashba}}+\mathcal{H}_{\mr{sd}},	\label{eq1-1} \\
	\mathcal{H}_{\mr{hop}}&=-t\sum_{(i,j),\sigma}(c^{\dagger}_{i\sigma}c_{j\sigma}+\mr{h.c.}), \label{eq1-2} \\
	\mathcal{H}_{\mr{Rashba}}&=\frac{E_R}{2}\sum_{j\sigma\sigma'}\left[ i c^{\dagger}_{j+\hat{y},\sigma}(\sigma_x)_{\sigma\sigma'}c_{j\sigma'} \right. \nn
 & \qquad \; \qquad \qquad \left. -ic^{\dagger}_{j+\hat{x}, \sigma}(\sigma_y)_{\sigma\sigma'}c_{j\sigma'}+\mr{h.c.}\right], \\
	\mathcal{H}_{\mr{sd}}&=\frac{J_{\mr{sd}}}{2} \sum_{j} \mbox{\boldmath{$S$}}_j \cdot c^{\dagger}_{j\sigma}(\mbox{\boldmath{$\sigma$}})_{\sigma\sigma'}c_{j\sigma'}, \label{eq1-4}
\end{align}
where $c_{i\sigma}$ is an annihilation operator of an electron with spin $\sigma$ at site $i$, $\sigma_{x(y)}$ is the $x$- ($y$-) component of Pauli matrices, $t$ is a hopping integral, $E_R$ is the Rashba energy, and $J_{\mr{sd}}$ is the sd coupling constant. The summation of $(i,j)$ represents the sum of the nearest-neighbor pairs and $\hat{x}$ $(\hat{y})$ is a unit vector in the $x$- ($y$-) direction. A schematic of the model is shown in Fig. \ref{f0}.
The sd interaction $\mathcal{H}_{\mr{sd}}$ is assumed between the magnetic moment $\mbox{\boldmath{$S$}}_j$ in the ferromagnet layer and the electron spin at the nearest-neighbor site in the heavy metal layer.

We calculate the RKKY interaction between magnetic moments $\mbox{\boldmath{$S$}}_j$ and $\mbox{\boldmath{$S$}}_{j+\textbf{r}}$, denoted $\mbox{\boldmath{$S$}}_1$ and $\mbox{\boldmath{$S$}}_2$, via the electrons represented by the Hamiltonian in eq. (\ref{eq1-1}). Here, $\mbox{\boldmath{$r$}}$ is the vector from $\mbox{\boldmath{$S$}}_1$ to $\mbox{\boldmath{$S$}}_2$. Generally speaking, the RKKY interaction is given by
\be 
	\mathcal{H}_{\mr{RKKY}}=\frac{J_{\mr{sd}}^2T}{8}\sum_m \mr{Tr}\left[(\mbox{\boldmath{$S$}}_1\cdot \mbox{\boldmath{$\sigma$}})G(\mbox{\boldmath{$r$}};i\omega_m)(\mbox{\boldmath{$S$}}_2\cdot \mbox{\boldmath{$\sigma$}})G(\mbox{-\boldmath{$r$}};i\omega_m)\right], \label{eq2}
\ee
where $\omega_m=\pi (2m+1)T$ is the Matsubara frequency with temperature $T$ and $G(\mbox{\boldmath{$r$}}; i\omega_m)$ is the $2\times 2$ Green function of the electron:
\begin{align}
	\qquad G(\mbox{\boldmath{$r$}}; i\omega_m)&=\sum_{\textbf{k}} \; G(\mbox{\boldmath{$k$}};i\omega_m)e^{i\textbf{k}\cdot \textbf{r}}, \\
	\qquad G(\mbox{\boldmath{$k$}};i\omega_m)&=\frac{1}{i\omega_m-\mathcal{H}_{\textbf{k}}+\mu}.\label{eq2-1}
\end{align}
Here, $\mathcal{H}_{\textbf{k}}$ is the Fourier transform of $\mathcal{H}_{\mr{hop}}+\mathcal{H}_{\mr{Rashba}}$ with wave vector $\mbox{\boldmath{$k$}}$ and $\mu$ is the chemical potential. $\mathcal{H}_{\textbf{k}}$ can be diagonalized, and the eigenenergies $\epsilon_{\pm \textbf{k}}$ are given by 
\be
\epsilon_{\pm\textbf{k}}=-2t(\mr{cos} k_x a+\mr{cos} k_y a)\pm E_R\sqrt{\mr{sin}^2 k_x a+\mr{sin}^2 k_y a}, \label{eq-eigen}
\ee
where $a$ is the lattice constant of the heavy metal.
Using the eigenstates $\left| \pm,\mbox{\boldmath{$k$}} \right \rangle$ corresponding to $\epsilon_{\pm\textbf{k}}$, we rewrite $\mathcal{H}_{\mr{RKKY}}$ in eq. (\ref{eq2}) as

\begin{align}
		\mathcal{H}_{\mr{RKKY}}&=\frac{J_{\mr{sd}}^2}{8}T\sum_m \sum_{n,n'=\pm}\sum_{\textbf{k},\textbf{q}}\sum_{\alpha,\beta}G_{n}(\mbox{\boldmath{$k$}}+\mbox{\boldmath{$q$}};i\omega_m)G_{n'}(\mbox{\boldmath{$k$}};i\omega_m) \nn
&\quad \times\langle n',\mbox{\boldmath{$k$}}| \sigma_{\alpha}| n,\mbox{\boldmath{$k$}}+\mbox{\boldmath{$q$}}\rangle
\langle n,\mbox{\boldmath{$k$}}+\mbox{\boldmath{$q$}}|\sigma_{\beta}| n',\mbox{\boldmath{$k$}} \rangle  e^{i\textbf{q}\cdot \textbf{r}} S_{1\alpha}S_{2\beta} \nn
&=\frac{J_{\mr{sd}}^2}{8}\sum_{n,n'=\pm}\sum_{\textbf{k},\textbf{q}}\sum_{\alpha,\beta} \frac{f(\epsilon_{n\textbf{k}+\textbf{q}})-f(\epsilon_{n'\textbf{k}})}{\epsilon_{n\textbf{k}+\textbf{q}}-\epsilon_{n'\textbf{k}}} \nn
&   \quad \times \langle n',\mbox{\boldmath{$k$}} | \sigma_{\alpha} | n,\mbox{\boldmath{$k$}}+\mbox{\boldmath{$q$}} \rangle \langle n,\mbox{\boldmath{$k$}}+\mbox{\boldmath{$q$}} | \sigma_{\beta}  | n',\mbox{\boldmath{$k$}} \rangle e^{i\textbf{q}\cdot \textbf{r}} S_{1\alpha}S_{2\beta}, \label{eq4}
\end{align}
where $G_{n}(\mbox{\boldmath{$k$}};i\omega_m)=1/(i\omega_m-\epsilon_{n\textbf{k}}+\mu)$ and $f(\epsilon)$ is the Fermi distribution function. When $\mbox{\boldmath{$r$}}$ is in the $x$-direction, we find that $\mathcal{H}_{\mr{RKKY}}$ can be expressed as
\be	\mathcal{H}_{\mr{RKKY}}=\sum_{\alpha=x,y,z}J_{\alpha}S_{1\alpha}S_{2\alpha}+D(S_{1z}S_{2x}-S_{1x}S_{2z}).
\label{eq-effint}
\ee
Other anisotropic interactions such as $\Gamma_{\alpha\beta} S_{1\alpha}S_{2\beta}$ ($\alpha\neq \beta$) vanish because of the symmetry requirement. The magnitude of the DMI vector $D$ is given by
\begin{align}
D&=D_++D_-+D_{\mr{inter}}, \label{eq3-1}\\
D_+&=\frac{J_{\mr{sd}}^2}{8}\sum_{\textbf{k},\textbf{q}} \frac{f(\epsilon_{+\textbf{k}+\textbf{q}})-f(\epsilon_{+\textbf{k}})}{\epsilon_{+\textbf{k}+\textbf{q}}-\epsilon_{+\textbf{k}}}e^{iq_xr} \nn
&  \qquad \qquad \times \langle +,\mbox{\boldmath{$k$}} | \sigma_z | +,\mbox{\boldmath{$k$}}+\mbox{\boldmath{$q$}} \rangle \langle +,\mbox{\boldmath{$k$}}+\mbox{\boldmath{$q$}} | \sigma_x  | +,\mbox{\boldmath{$k$}} \rangle, \label{eq3-2}\\
D_-&=\frac{J_{\mr{sd}}^2}{8}\sum_{\textbf{k},\textbf{q}} \frac{f(\epsilon_{-\textbf{k}+\textbf{q}})-f(\epsilon_{-\textbf{k}})}{\epsilon_{-\textbf{k}+\textbf{q}}-\epsilon_{-\textbf{k}}}e^{iq_xr} \nn
&  \qquad \qquad \times \langle -,\mbox{\boldmath{$k$}} | \sigma_z | -,\mbox{\boldmath{$k$}}+\mbox{\boldmath{$q$}} \rangle \langle -,\mbox{\boldmath{$k$}}+\mbox{\boldmath{$q$}} | \sigma_x  | -,\mbox{\boldmath{$k$}} \rangle, \label{eq3-3}\\
D_{\mr{inter}}&=\frac{J_{\mr{sd}}^2}{4}\sum_{\textbf{k},\textbf{q}}\frac{f(\epsilon_{+\textbf{k}+\textbf{q}})-f(\epsilon_{-\textbf{k}})}{\epsilon_{+\textbf{k}+\textbf{q}}-\epsilon_{-\textbf{k}}}e^{iq_xr} \nn
&  \qquad \qquad \times \langle -,\mbox{\boldmath{$k$}} | \sigma_z | +,\mbox{\boldmath{$k$}}+\mbox{\boldmath{$q$}} \rangle \langle +,\mbox{\boldmath{$k$}}+\mbox{\boldmath{$q$}} | \sigma_x  | -,\mbox{\boldmath{$k$}} \rangle. \label{eq3-4}
\end{align}
where $D_+$ ($D_-$) represents the intraband contribution of the $\epsilon_{+\textbf{k}}$ ($\epsilon_{-\textbf{k}}$) band, whereas $D_{\mr{inter}}$ is the interband contribution.

First, let us discuss the direction of the $\mbox{\boldmath{$D$}}$ vector. Equation (\ref{eq-effint}) indicates that the DMI vector $\mbox{\boldmath{$D$}}$ is parallel to the $y$-direction. This is consistent with the direction determined from the following symmetry argument. Since $\mbox{\boldmath{$S$}}_1$ and $\mbox{\boldmath{$S$}}_2$ are aligned in the $x$-direction, there are two planes of inversion symmetry. 
One is the $yz$-plane passing through the midpoint of the two magnetic moments. 
The other is the $zx$-plane passing through the two magnetic moments. 
Owing to these planes of inversion symmetry, the DMI vector $\mbox{\boldmath{$D$}}$ should be parallel to the $yz$-plane and perpendicular to the $zx$-plane so that $\mbox{\boldmath{$D$}}$ is parallel to the $y$-direction. \cite{crepi}

\section{Numerical Results}
\begin{figure}[h]
\includegraphics[width=7cm]{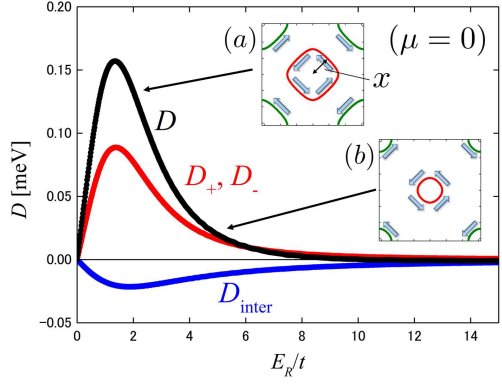}
\centering
\caption{(Color online) Rashba energy dependences of $D$ (black), $D_{+}$, $D_{-}$ (red), and $D_{\mr{inter}}$ (blue) for $t=0.3 \mr{eV}$, $J_{\mr{sd}}=0.1 \mr{eV}$, $T=300 \mr{K}$, $\mu=0$, and $r=a$. The insets show the configuration of the Fermi surfaces of both $\epsilon_{+\textbf{k}}$ (red) and $\epsilon_{-\textbf{k}}$ (green) in the first Brillouin zone for (a) $E_R=2t$ and (b) $E_R=5t$ ($\mu=0$). 
In insets (a) and (b), the spin texture caused by the Rashba term Eq. (3) at the Fermi surfaces is depicted by the arrows.}
\label{f1}
\end{figure}

Figure \ref{f1} shows the Rashba energy dependence of the DMI for the case of $\mu=0$ and $r=a$. Note that $D$ is an odd function of $E_R/t$ and is proportional to $J_{\mr{sd}}^2/t$. In order to see the energy scale, we choose $t=0.3 \mr{eV}$, $J_{\mr{sd}}=0.1 \mr{eV}$, and $T=300 \mr{K}$ as typical values.
As shown in Fig. \ref{f1}, $D$ increases linearly for $E_R <t$, has a maximum near $E_R \simeq 1.3t$, and decreases for $E_R >2t$. The maximum value of $D$ is around $0.16 \mr{meV}$. $D_+$, $D_-$, and $D_{\mr{inter}}$ have similar behavior except that $D_{\mr{inter}}$ has the opposite sign.
The temperature dependence of the DMI is represented by the Fermi distribution function in Eqs. (12)-(14). The temperature dependence is small.

The linear dependence of the DMI on $E_R/t$ in the small-$E_R$ region is similar to that calculated in a two-dimensional electron gas \cite{imamura}. On the other hand, the behavior in the large-$E_R$ region is considerably different from that in the two-dimensional electron gas; the DMI oscillates without decreasing in the large $E_R$ region. In contrast, our results show that the DMI decreases and approaches zero for large $E_R$.
In order to see the reason for this difference, we study the $E_R$ dependence of the Fermi surface.
As shown in the inset of Fig. \ref{f1}, we first define $x$ as the diagonal distance from the origin in the first Brillouin zone to the Fermi surface. $x$ is related to $E_R/t$ by the relation $\mr{tan}(\pi/2-x/\sqrt{2})=E_R/(2\sqrt{2}t)$. When $E_R$ is large, $x \simeq 4t/E_R$.
In Eqs. (\ref{eq3-2})-(\ref{eq3-4}), the terms $[f(\epsilon_{n\textbf{k}+\textbf{q}})-f(\epsilon_{n'\textbf{k}})]/(\epsilon_{n\textbf{k}+\textbf{q}}-\epsilon_{n'\textbf{k}})$ ($n, n'=\pm$) have dominant values on the Fermi surface, whose length is approximately proportional to $x$. The matrix elements do not depend on $E_R$. 
Therefore, when $E_R$ is large, the contribution to $D$ is proportional to $x \simeq 4t/E_R$ and approaches zero for large $E_R$.

\begin{figure}[h]
\includegraphics[width=7cm]{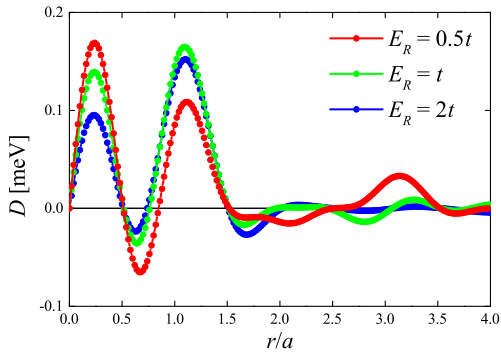}
\centering
\caption{(Color online) Spin distance dependences of DMI for $E_R=0.5t$, $t$, and $2t$ ($\mu=0$).}
\label{f3}
\end{figure}

Next we study the $\mbox{\boldmath{$r$}}$ dependence of the DMI, which is shown in Fig. \ref{f3}. We find that its sign changes as the spin distance $r$ increases. The DMI is maximum at around $r/a \simeq 1.1$ or $0.24$. The origin of this oscillation is the same as that of the usual RKKY interaction. However, the oscillation amplitude does not always decrease monotonically, which is different from the usual RKKY interaction in the two-dimensional electron gas. 

\begin{figure}[h]
\includegraphics[width=7cm]{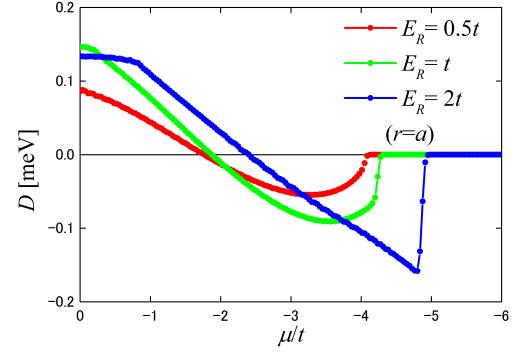}
\centering
\caption{(Color online) Chemical potential dependences of the DMI for $E_R=0.5t$, $t$, and $2t$ ($r=a$).}
\label{f4}
\end{figure}

The dependence of the DMI on the chemical potential is shown in Fig. \ref{f4}. The sign of $D$ depends on the chemical potential. 
It has a maximum when the heavy metal is at half-filling ($\mu \simeq 0$). 
As $E_R$ increases, $D$ becomes flat around $\mu \simeq 0$. 

\begin{figure}[h]
\includegraphics[width=7cm]{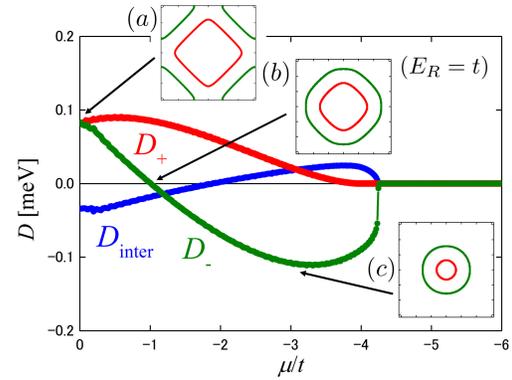}
\centering
\caption{(Color online) Chemical potential dependences of $D_+$ (red), $D_-$ (green), and $D_{\mr{inter}}$ (blue) for $E_R=t$ and $r=a$. The insets show the configuration of the Fermi surfaces of both $\epsilon_{+\textbf{k}}$ (red) and $\epsilon_{-\textbf{k}}$ (green) in the first Brillouin zone for (a) $\mu=0$, (b) $\mu=-t$, and (c) $\mu=-3t$ ($E_R=t$).}
\label{f5}
\end{figure}

\begin{figure}[h]
\includegraphics[width=7cm]{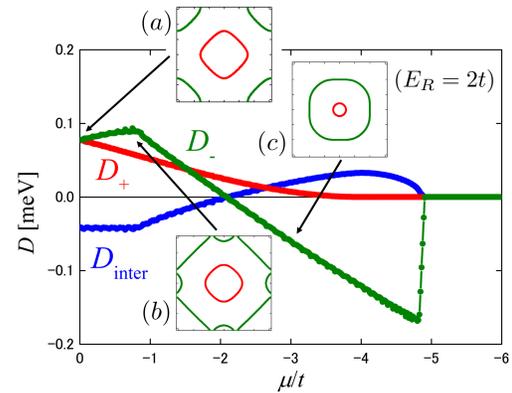}
\centering
\caption{(Color online) Same as in Fig. \ref{f5} for $E_R=2t$. The insets show the configuration of Fermi surfaces for (a) $\mu=0$, (b) $\mu=-0.8t$, and (c) $\mu=-3t$ ($E_R=2t$).}
\label{f6}
\end{figure}

In order to understand this chemical potential dependence, we show the contributions of $D_+$, $D_-$, and $D_{\mr{inter}}$ in Figs. \ref{f5} and \ref{f6} for $E_R=t$ and $2t$, respectively. 
The Fermi surface of the $\epsilon_{+\textbf{k}}$ band (red) (shown in the insets) monotonically shrinks as $|\mu|$ increases and disappears when $\mu \leq -4t$. Therefore, $D_+$ decreases almost monotonically. The Fermi surface of the $\epsilon_{-\textbf{k}}$ band (green) also shrinks, but $D_-$ changes its sign around the point where the dominant nesting vector satisfies $q_xa \simeq \pi$. Since the matrix element in Eq. (\ref{eq3-3}) is purely imaginary, only $i\mr{sin}q_xr$ of $\mr{exp}iq_xr$ remains ($r=a$). Therefore, when $q_xa$ crosses $\pi$, Eq. (\ref{eq3-3}) changes sign. 

Next, we consider the case of $E_R=2t$ in order to see the reason why $D$ is flat around $\mu \simeq 0$ in Fig. \ref{f3}. It is found in Fig. \ref{f6} that $D_-$ has a good nesting around $\mu \simeq -0.8t$ and has a maximum. From this effect, 
$D$ becomes flat in the range $0 \geq \mu \gtrsim -0.8t$.

\section{Comparison with Experiments}
Finally, we discuss the implications of the present results for experiments. The Rashba energy has been estimated as $0.1-1 \mr{eV}$ \cite{miron2, suzuki} from the measurement of the transverse magnetic field induced by the Rashba effect at the interfaces of ferromagnet/heavy metal heterostructures. This corresponds to $E_R/t=0.3-3.0$. From Fig. \ref{f1}, the value of the DMI is on the order of $0.1 \mr{meV}$. Therefore, for a ferromagnet whose lattice constant is $3 \AA$ \cite{burton}, the DMI per square area is on the order of $0.1 \mr{mJ/m^2}$ which is comparable to some experimental data \cite{emori, torrejon}. 

The sign change of the DMI has been observed experimentally for heavy metals whose $r/a$ is in the range of $0.7-0.9$ \cite{torrejon}. This range is close to the range shown in Fig. \ref{f3} where the sign change occurs. Therefore, we expect that the main origin of the sign change is the oscillation of the RKKY interaction. We also expect that the chemical potential can be an additional origin of the sign change.

As discussed in the introduction, it is argued that the domain wall in the ferromagnet is driven by the spin torque induced by the spin Hall effect and the Rashba effect when current flows in the heavy metal \cite{miron2, liu2, kimras, khval}. In this mechanism, the DMI due to the Rashba SOI helps the domain wall motion, \textit{i.e.}, enhances the velocity of the domain wall \cite{thiaville}.  
However, as shown in Fig. \ref{f1}, DMI becomes small when $E_R$ is large. Therefore, a large Rashba SOI does not always induce a high velocity of the domain wall motion \cite{shibuya}.

As shown in Fig. \ref{f00}, there are two mechanisms and DMI can also be induced by the SOI in ferromagnets near the interface. Freimuth \textit{et} \textit{al}. estimated the DMI by the Berry phase formulation~\cite{Freimuth}. However, the obtained DMI cannot be divided into the components of the two mechanisms. It remains a future problem to evaluate the strength of DMI caused by these two mechanisms under actual experimental situations.

\section{Conclusions}
We studied a microscopic derivation and found several properties of the DMI between magnetic moments in a ferromagnet/heavy metal heterostructure. We derived the DMI on the basis of the RKKY interaction through electrons in a heavy metal with Rashba SOI. 
The dependence of the DMI on the Rashba energy is different from that obtained in a two-dimensional electron gas. Furthermore, we found that the sign change of DMI occurs as a function of distance between the two magnetic moments as well as a function of the chemical potential of the heavy metal.
The obtained results will be useful for studying materials that provide large DMI. 

One of the authors (H. M.) thanks T. Ono for useful discussions. This work was supported by the JSPS Core-to-Core Program "Advanced Research Networks", and a Grant-in-Aid for Scientific Research on Innovative Areas "Ultra Slow Muon Microscope" (No. 23108004) from the Ministry of Education, Culture, Sports, Science and Technology, Japan. We were also supported by Grants-in-Aid for Scientific Research from the Japan Society for the Promotion of Science (Nos. 15K17694, 25220803, and 15H02108).

\end{document}